\documentclass[pra,twocolumn]{revtex4}
\usepackage{graphicx}

\begin{document}
\title{Generation of entangled squeezed states in atomic Bose-Einstein condensates}
\author{Le-Man Kuang\footnote{Corresponding
author.}\footnote{Email address: lmkuang@hunnu.edu.cn (L. M.
Kuang)}$^{1,2}$, Ai-Hua Zeng$^{3}$ and Zhen-Hua Kuang$^{3}$}
\address{$^{1}$Department of Physics, Hunan Normal University, Changsha 410081, People's Republic of China\\
$^{2}$The Abdus Salam International Centre for Theoretical
Physics, Strada Costiera 11, Trieste 34014, Italy\\
$^{3}$Department of Physics, Shaoyang University, Shaoyang 422000,
People's Republic of China }
\begin{abstract}
A method for producing entangled squeezed states (ESSs) for atomic
Bose-Einstein condensates (BECs) is proposed by using a BEC with
three internal states and two classical laser beams. We show that
it is possible to generate two-state and multi-state ESSs under
certain circumstances.

\noindent PACS number(s): 03.75.Fi, 03.65.Ud, 03.65.Ta, 42.50.Dv
\end{abstract}

\maketitle

\section{Introduction}
Quantum entanglement has been the focus of much work in the
foundations of quantum mechanics, being particularly with quantum
nonseparability, the violation of Bell's inequalities, and the
so-called Einstein-Pololsky-Rosen (EPR) paradox. Beyond this
fundamental aspect, creating and manipulating of entangled states
are essential for quantum information applications. Among these
applications are quantum computation \cite{nie}, quantum
teleportation \cite{ben1}, quantum dense coding \cite{ben2}, and
quantum cryptography \cite{ben3}. Hence, quantum entanglement has
been viewed as an essential resource for quantum information
processing.

In recent years, much progress has been made on creating quantum
entanglement between macroscopic atomic samples
\cite{dua1,jul,sor,dua2,dua3,hel,pu,dua4}. There are several
proposals to generate quantum entanglement between macroscopic
atomic ensembles \cite{dua3} and to explore its applications to
quantum communication \cite{dua1,dua0,kuz} and quantum computation
\cite{you}. In particular, quantum entanglement between two
separate macroscopic atomic samples  \cite{jul} has been
demonstrated experimentally. On the aspect of atomic Bose-Einstein
condensates (BECs) it has been shown that substantial
many-particle entanglement can be generated directly in a
two-component weakly interacting BEC using the inherent
inter-atomic interactions \cite{sor,sor1} and a spinor BEC using
spin-exchange collision interactions \cite{dua2,pu,dua4}. Based on
an effective interaction between two atoms from coherent Raman
processes, Helmerson and You \cite{hel} proposed a coherent
coupling scheme to create massive entanglement of BEC atoms.  An
entanglement swapping scheme between trapped BECs \cite{dun} has
also been proposed. Indeed, nowadays manipulation and control of
quantum entanglement between BEC atoms has become one of important
goals for experimental studies with BECs. As well known, one of
the key problems in the experimental explorations of quantum
entanglement is to coherently control interaction between the
relevant particles. The strength of the inter-atomic interactions
in atomic BECs can vary over a wide range of values through
changing external fields. This kind of control and manipulation of
inter-atomic interactions has been experimentally realized through
magnetical-field-induced Feshbach resonances in atomic BECs
\cite{ino}. Therefore, atomic BECs provide us with an ideal
experimental system for studying quantum entanglement.

On the other hand, recently much attention has been paid to
continuous variable quantum information processing in which
continuous-variable-type entangled pure states play a key role.
For instance, two-state entangled coherent states are used to
realize efficient quantum computation \cite{jeo} and quantum
teleportation \cite{enk}. Two-mode squeezed vacuum states have
been applied to quantum dense coding \cite{ban}. In particular,
following the theoretical proposal of Ref. \cite{bra}, continuous
variable teleportation has been experimentally demonstrated  for
coherent states of a light field \cite{fur} by using entangled
two-mode squeezed vacuum states produced by parametric
down-conversion in a sub-threshold optical parametric oscillator.
It is also has been shown that a two-state entangled squeezed
vacuum state can be optically created and used to realize quantum
teleportation of an arbitrary coherent superposition state of two
equal-amplitude and opposite-phase squeezed vacuum states
\cite{zho,cai}. Therefore, it is an interesting topic to create
entangled squeezed states in atomic BECs.

In this paper, we present a scheme to produce entangled squeezed
states  for atomic BECs. The proposed system consists of an atomic
BEC with three internal states and two classical laser beams with
appropriate frequencies. They form a three-level lambda
configuration. We show that it is possible to generate entangled
squeezed states for atomic BECs. This paper is organized as
follows. In Sec. II, we present the physical system under our
consideration, establish our model, and give an approximate
analytic solution of the model. In Sec. III, we show how to
produce entangled squeezed vacuum states for atomic BECs. We shall
conclude our paper with discussions and remarks in the last
section.

\section{Model and solution}

Consider a cloud of BEC atoms which have three internal states
labelled by $|1\rangle$,  $|2\rangle$, and  $|3\rangle$ with
energies $E_1$,   $E_2$, and $E_3$, respectively. The two lower
states   $|1\rangle$ and  $|3\rangle$  are Raman coupled to the
upper state  $|2\rangle$ via, respectively, two classical laser
fields of frequencies $\omega_1$ and $\omega_2$  in the Lambda
configuration. The interaction scheme is shown in Fig. 1. The
atoms in these internal states are subject to isotropic harmonic
trapping potentials $V_i({\bf r})$ for $i=1,2,3$, respectively.
Furthermore, the atoms in BEC interact with each other via elastic
two-body collisions with the $\delta$-function potentials
$V_{ij}({\bf r}-{\bf r}')=U_{ij}\delta ({\bf r}-{\bf r}')$,where
$U_{ij}=4\pi\hbar^2a_{ij}/m$ with $m$ and  $a_{ij}$, respectively,
being the atomic mass and  the $s$-wave scattering length between
atoms in states $i$ and $j$. A good experimental candidate of this
system is the sodium atom condensate for which there exist
appropriate atomic internal levels and external laser fields to
form the Lambda configuration which has been used to demonstrate
ultraslow light propagation \cite{hau} and amplification of light
and atoms \cite{ino1} in atomic BECs.

The second quantized Hamiltonian to describe the system at zero
temperature is given by
\begin{equation}
\label{1}
\hat{H}=\hat{H}_{a}+\hat{H}_{a-l}+\hat{H}_{c},
\end{equation}
where $\hat{H}_{a}$ gives the free evolution of the atomic fields,
$\hat{H}_{a-l}$ describes the dipole interactions between the atomic fields and laser fields,
and $\hat{H}_{c}$ represents inter-atom two-body interactions.

The free atomic Hamiltonian is given by
\begin{equation}
\label{2}
\hat{H}_{a}=\sum^3_{i=0}\int d{\bf x} \hat{\psi}^{\dagger}_i({\bf x})
\left [-\frac{\hbar^2}{2m}\nabla^2 +V_i({\bf x})+E_i\right ]
\hat{\psi}_i({\bf x}),
\end{equation}
where $E_i$ are internal energies for the three internal states,
$\hat{\psi}_i({\bf x})$ and $\hat{\psi}^{\dagger}_i({\bf x})$ are
the boson  annihilation  and creation operators for the
$|i\rangle$-state atoms at position ${\bf x}$, respectively, they
satisfy   the standard boson commutation relation
$[\hat{\psi}_i({\bf x}), \hat{\psi}^{\dagger}_j({\bf
x}')]=\delta_{ij}\delta({\bf x}-{\bf x}')$ and $[\hat{\psi}_i({\bf
x}), \hat{\psi}_j({\bf x}')] =0=[\hat{\psi}^{\dagger}_i({\bf x}),
\hat{\psi}^{\dagger}_j({\bf x}')]$.

The atom-laser interactions in the dipole approximation can be described by the following Hamiltonian
\begin{eqnarray}
\label{3}
\hat{H}_{a-l}&=&\frac{1}{2} \int d{\bf x}  \left [\Omega_1
 \hat{\psi}^{\dagger}_2({\bf x})\hat{\psi}_1({\bf x})
e^{i({\bf k_1}\cdot {\bf x}-\omega_1t)} \right. \nonumber \\
& &\left.+\Omega_2\hat{\psi}^{\dagger}_2({\bf x})
\hat{\psi}_3({\bf x})e^{i({\bf k_2}\cdot {\bf x}-\omega_2t)}+H.c.\right ],
\end{eqnarray}
where $\Omega_1=-\mu_{21}{\cal E}_1/\hbar$ and
$\Omega_2=-\mu_{23}{\cal E}_2/\hbar$ are the Rabi frequencies of
the two laser beams with $\mu_{ij}$ denoting   a  transition
dipole-matrix element between states $|i\rangle$ and $|j\rangle$,
${\bf k}_1$ and ${\bf k}_2$ are wave vectors of correspondent
laser fields.

The collision Hamiltonian is taken to be the following form
\begin{eqnarray}
\label{4}
\hat{H}_{c}&=&\frac{2\pi\hbar^2}{m}
\int d{\bf x}\left [\sum^3_{i=1}a^{sc}_i\hat{\psi}^{\dagger}_i({\bf x})
\hat{\psi}^{\dagger}_i({\bf x})\hat{\psi}_i({\bf x})\hat{\psi}_i({\bf x})
\right. \nonumber \\
& &\left.+\sum_{i\neq j}2a^{sc}_{ij} \hat{\psi}^{\dagger}_i({\bf x})
\hat{\psi}^{\dagger}_j({\bf x})\hat{\psi}_i({\bf x})\hat{\psi}_j({\bf x})
\right ],
\end{eqnarray}
where $a^{sc}_i$ is  $s$-wave scattering length of condensate in
the internal state $|i\rangle$ and  $a^{sc}_{ij}$ that between
condensates in the internal states $|i\rangle$ and $|j\rangle$.

\begin{figure}[htb]
\begin{center}
\includegraphics[width=7cm]{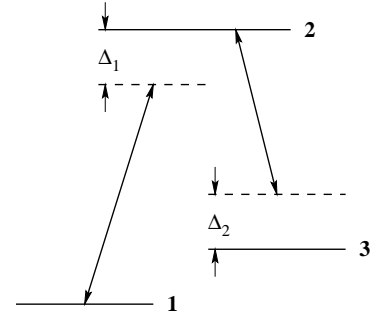}
\end{center}
\vskip 0.2cm \caption{Three-level Lambda-type atoms coupled to two
classical laser fields with the detunings $\Delta_1$ and
$\Delta_2$.} \label{fig1}
\end{figure}

For a weakly interacting BEC at zero temperature one may neglect
all modes except for the condensate mode and approximately
factorize the atomic field operators as a product of a single mode
operator $\hat{b}_i$ and  a normalized wavefunction for the atoms
in the BEC $\phi_i({\bf x})$, i.e.,   $\hat{\psi}_i({\bf
x})\approx \hat{b}_i\phi_i({\bf x})$ where $\phi_i({\bf x})$ is
given by the ground state of the following Schr\"{o}dinger
equation
\begin{equation}
\label{4'} \left [-\frac{\hbar^2}{2m}\nabla^2 +V_i({\bf
x})+E_i\right ] {\phi}_i({\bf x})=\hbar\nu_i{\phi}_i({\bf x}),
\end{equation}
where $\hbar\nu_i$ is the energy of the mode $i$.

The valid conditions of the single-mode approximation were
demonstrated in Refs. \cite{milb,kuan}, which indicate that this
approximation provides a reasonably accurate picture for weak
many-body interactions, i.e., for  small number of condensed
atoms.  For large condensates, the mode functions of condensates
are altered due to the  collision  interactions, and the two-mode
approximation breaks down.  A simple estimate shows that this
happens when the number of atoms $N$ satisfies $Na^{sc}\gg r_0$,
where $a^{sc}$ is a typical scattering length and $r_0$ is  a
measure of the trap size. If we consider a large trap \cite{hau}
with the size $r_0=100 \mu$m and  the  typical scattering length
$a^{sc}=5$ nm, the single mode approximation is applicable for
$N\le 20000$. Substituting the single-mode expansions of the
atomic field operators into Eqs. (\ref{2}-\ref{4}), we arrive at
the following three-mode approximate Hamiltonian
\begin{eqnarray}
\label{5}
\hat{H}&=&\hbar\sum^3_{i=1}\nu_i\hat{b}^{\dagger}_i\hat{b}_i
-\hbar\left [g_1\hat{b}^{\dagger}_2\hat{b}_1e^{-i\omega_1t} \right.
\nonumber \\
&&\left. +g_2\hat{b}^{\dagger}_2\hat{b}_3e^{-i\omega_2t}+H.c. \right ]
\nonumber \\
&&+\sum^3_{i=1}\lambda_i\hat{b}^{\dagger 2}_i\hat{b}^2_i
+\sum_{i\neq j}\lambda_{ij}\hat{b}^{\dagger}_i\hat{b}_i
\hat{b}^{\dagger}_j\hat{b}_j,
\end{eqnarray}
where  $g_i$ are the linear  coupling constants  defined by
\begin{equation}
\label{7}
  g_i=\frac{1}{2}\Omega_i\int d{\bf x} \phi^*_{2}({\bf x})\phi_{1}({\bf x})e^{i{\bf k}_i.{\bf
  x}}.
\end{equation}
 And  $\lambda_{i}$ and $\lambda_{ij}$ are nonlinear coupling
 constants given by
\begin{eqnarray}
\label{8}
  \lambda_i&=& \frac{2\pi\hbar^2a^{sc}_i}{m}\int d{\bf x}|\phi_{i}({\bf x})|^4,\\
\label{9}
  \lambda_{ij}&=&\frac{4\pi\hbar^2a^{sc}_{ij}}{m}\int d{\bf x}
|\phi_{i}({\bf x})|^2|\phi_{j}({\bf x})|^2,
 \hspace{0.3cm}(i\neq j).
\end{eqnarray}

 Going over to an interaction picture with respect to
\begin{equation}
\label{10}
H_0=\hbar\nu_1\sum^3_{i=1}\hat{b}^{\dagger}_i\hat{b}_i+\hbar(\omega_1-\omega_2)\hat{b}^{\dagger}_3\hat{b}_3
+\hbar\omega_1\hat{b}^{\dagger}_2\hat{b}_2,
\end{equation}
 we can transfer the time-dependent Hamiltonian (\ref{5}) to the following
 time-independent Hamiltonian
\begin{eqnarray}
\label{11}
 \hat{H}_I &=&\hbar(\Delta_1-\Delta_2)\hat{b}^{\dagger}_3\hat{b}_3+\hbar\Delta_1\hat{b}^{\dagger}_2\hat{b}_2
 \nonumber \\
& & -\hbar[g_1\hat{b}^{\dagger}_2\hat{b}_1+
g_2\hat{b}^{\dagger}_2\hat{b}_3+H.c.]
\nonumber \\
& & +\sum^3_{i=1}\lambda_i\hat{b}^{\dagger 2}_i\hat{b}^2_i
+\sum_{i\neq j}\lambda_{ij}\hat{b}^{\dagger}_i\hat{b}_i
\hat{b}^{\dagger}_j\hat{b}_j,
\end{eqnarray}
where $\Delta_1=\nu_2-\nu_1-\omega_1$ and $\Delta_2=\nu_2-\nu_3-\omega_2$ are the detunings of the two laser beams,
respectively.

We consider the situation of the exact two-photon resonance (i.e.,
$\Delta_1=\Delta_2=\Delta$), and suppose the large two-photon
detuning $\Delta\gg \nu_3-\nu_1$. In this case  from the
Hamiltonian (\ref{10}) the atomic field operators $\hat{b}_2$ and
$\hat{b}^{\dagger}_2$  can be adiabatically eliminated. Then  we
arrive at the following effective two-mode Hamiltonian containing
only atomic field operators in internal states $|1\rangle$ and
$|3\rangle$
\begin{eqnarray}
\label{13} \hat{H}_{eff}&=&\omega_1\hat{b}^{\dagger}_1\hat{b}_1
+\omega_3 \hat{b}^{\dagger}_3\hat{b}_3 +
(g\hat{b}^{\dagger}_3\hat{b}_1
+ g^*\hat{b}^{\dagger}_1\hat{b}_3)\nonumber \\
& &+\lambda_1\hat{b}^{\dagger 2}_1\hat{b}^2_1
+\lambda_{13}\hat{b}^{\dagger}_1\hat{b}_1\hat{b}^{\dagger}_3\hat{b}_3
+ \lambda_3\hat{b}^{\dagger 2}_3\hat{b}^2_3,
\end{eqnarray}
where we have set $\hbar=1$ and introduced
\begin{equation}
\label{14}
\omega_1=-\frac{|g_1|^2}{\Delta}, \hspace{0.3cm}
\omega_3=-\frac{|g_2|^2}{\Delta}, \hspace{0.3cm}
g=-\frac{g_1g^*_2}{\Delta}.
\end{equation}

From Eqs. (\ref{11}) and (\ref{13}) we see that the laser-atom
interactions are converted as an atomic effective tunnelling
interaction between state $|1\rangle$ and state $|3\rangle$ with
the tunnelling coupling strength being determined by strengths of
the laser-atom interactions and the detuning. In the derivation of
Eq. (\ref{13}), all terms involving $\hat{b}^{\dagger}_2\hat{b}_2$
have been ignored since the atomic population in the internal
state $|2\rangle$ approaches zero under conditions of our
consideration.

For the sake of simplicity, we consider a symmetric interaction situation
in which inter-atomic interactions in condensates in the internal states
$|1\rangle$ and $3\rangle$ have the same interacting strengths and two applied
lasers have the same Rabi frequencies. So that we have $g_1=g_2$ and $\lambda_1=\lambda_3\equiv q$.
Then from Eq. (\ref{14}) we can obtain $\omega_1=\omega_3=-|g_1|^2/\Delta\equiv g$.
Hence the effective Hamiltonian (\ref{13}) reduces to the following simple form
\begin{eqnarray}
\label{15}
 \hat{H}_{eff} &=&g(\hat{b}^{\dagger}_1\hat{b}_1 + \hat{b}^{\dagger}_3\hat{b}_3)
+  q(\hat{b}^{\dagger 2}_1\hat{b}^2_1 + \hat{b}^{\dagger 2}_3\hat{b}^2_3)\nonumber \\
&&+  g(\hat{b}^{\dagger}_1\hat{b}_3 + \hat{b}^{\dagger}_3\hat{b}_1) + 2\chi\hat{b}^{\dagger}_1\hat{b}_1
\hat{b}^{\dagger}_3\hat{b}_3,
\end{eqnarray}
where we have set $\chi=\lambda_{13}/2$.
When $q$ and $\chi$ are much less than $|g|$, which is the case of weak inter-atomic
nonlinear interactions, the effective Hamiltonian can be solved approximately under the rotating-wave approximation.
In order to do this, one introduces the following unitary transformation
\begin{equation}
\label{16} \hat{b}_1=\frac{1}{\sqrt{2}}(\hat{B}_1
-i\hat{B}_3),\hspace{0.5cm}
\hat{b}_3=\frac{1}{\sqrt{2}}(\hat{B}_1 +i\hat{B}_3),
\end{equation}
 where $\hat{B}_1$ and $\hat{B}_3$ satisfy the usual boson commutation relations:
 $[\hat{B}_i, \hat{B}_j]=0=[\hat{B}^{\dagger}_i, \hat{B}^{\dagger}_j] $,
and   $[\hat{B}_i, \hat{B}^{\dagger}_j]=\delta_{ij}$
with $\hat{B}^{\dagger}_i$ being
 the hermitian conjugation of $\hat{B}_j$.
Under the rotating-wave approximation \cite{aga}, we get the following approximate Hamiltonian
\begin{eqnarray}
\label{17}
 \hat{H}_{eff} & \approx &\omega\hat{N} + g(\hat{B}^{\dagger}_1\hat{B}_1-\hat{B}^{\dagger}_3\hat{B}_3)
 \nonumber \\&&+\frac{1}{4}q[3\hat{N}^2 -(\hat{B}^{\dagger}_1\hat{B}_1-\hat{B}^{\dagger}_3\hat{B}_3)^2]
 \nonumber \\ &&+\frac{1}{2}\chi\hat{N}^2-\chi\hat{B}^{\dagger}_1\hat{B}_1\hat{B}^{\dagger}_3\hat{B}_3,
\end{eqnarray}
 where the total number operator $\hat{N}$ is a conserved constant which is given by
 $ \hat{N}=\hat{b}^{\dagger}_1\hat{b}_1+\hat{b}^{\dagger}_3\hat{b}_3
=\hat{B}^{\dagger}_1\hat{B}_1+\hat{B}^{\dagger}_3\hat{B}_3$, and we have introduced a new parameter
\begin{equation}
\label{18}
\omega=g- \frac{1}{2}(\chi+q).
\end{equation}

The bases of the Fock spaces in the  $(\hat{b}_1, \hat{b}_3)$ and
 $(\hat{B}_1, \hat{B}_3)$ representations are defined, respectively, by
\begin{eqnarray}
\label{19}
|n,m\rangle &=&\frac{1}{\sqrt{n!m!}}\hat{b}^{\dagger n}_1\hat{b}^{\dagger m}_3|0,0\rangle ,  \\
\label{20}
 |n,m)&=&\frac{1}{\sqrt{n!m!}}\hat{B}^{\dagger n}_1\hat{B}^{\dagger m}_3|0,0),
\end{eqnarray}
where $n$ and $m$ take non-negative integers.  Obviously,
$\hat{H}_{eff}$ is diagonal in the Fock space of  $(\hat{B}_1,
\hat{B}_3)$, and we have
 \begin{equation}
\label{21} \hat{H}_{eff}|n,m)=E(n,m)|n,m),
\end{equation}
where eigenvalues of the Hamiltonian  are given by the following expression
\begin{eqnarray}
\label{22}
 E(n,m)&=& \omega (n+m)+g(n-m)
  +\frac{1}{2}(q+\chi)(n+m)^2 \nonumber  \\
&& + (q -\chi) nm.
\end{eqnarray}

\section{Entangled squeezed  states}

In this section we shall show that entangled squeezed vacuum
states for atomic BECs can be produced when atomic BECs are
initially in a product squeezed vacuum state through properly
manipulating  laser-atom interactions and inter-atomic
interactions in the BECs.

Consider a product squeezed vacuum state of two squeezed vacuum states defined in Fock spaces of
$(\hat{b}_1, \hat{b}_3)$ and $(\hat{B}_1, \hat{B}_3)$,
 respectively,
\begin{eqnarray}
\label{53}
|\xi_1, \xi_3\rangle
&=& \hat{S}_{\hat{b}_1}(\xi_1)\hat{S}_{\hat{b}_3}(\xi_3) |0,0\rangle, \\
\label{54}
 |\eta_1, \eta_3)
&=& \hat{S}_{\hat{B}_1}(\eta_1)\hat{S}_{\hat{B}_3}(\eta_3) |0,0),
\end{eqnarray}
where the single mode squeezing operators in the $(\hat{b}_1,
\hat{b}_3)$ and $( \hat{B}_1, \hat{B}_3)$ representations with
arbitrary complex squeezing parameters $\xi_i$ and $\eta_i$
($i=1,3$) are defined by
\begin{eqnarray}
\label{53'} \hat{S}_{\hat{b}_i}(\xi_i)&=&
\exp\left[-\frac{1}{2}\left(\xi_i\hat{b}^{\dagger}_i -
\xi^*_i\hat{b}_i\right)\right], \\
\label{53''}
\hat{S}_{\hat{B}_i}(\eta_i)&=&\exp\left[-\frac{1}{2}\left(\eta_i
\hat{B}^{\dagger}_i - \eta^*_i \hat{B}_i\right)\right].
\end{eqnarray}

For the convenience in later use we here introduce a two-mode
squeezed state in the $(\hat{B}_1, \hat{B}_3)$ representation
\begin{equation}
\label{55}
|\zeta )_{B_1 B_3}= \hat{S}_{\hat{B}_1\hat{B}_3}(\zeta)|0,0),
\end{equation}
where the two-mode squeezing operator is defined by
\begin{equation}
\label{55'} \hat{S}_{\hat{B}_1\hat{B}_3}(\zeta) =
\exp\left(-\zeta\hat{B}^{\dagger}_1\hat{B}^{\dagger}_3
+\zeta^*\hat{B}_1\hat{B}_3\right),
\end{equation}
where  $\zeta$ is an arbitrary complex number.

It is straightforward to see that a direct-product state of two
squeezed vacuum states in the $(\hat{b}_1, \hat{b}_3)$ and
$(\hat{B}_1, \hat{B}_3)$ representations is transferred to an
entangled state in the  correspondent representation,
respectively. In general, the entangled state in correspondent
representation cannot be explicitly expressed as a product
squeezed vacuum state for general squeezing parameters $\xi$ and
$\eta$. However, a product squeezed state of two squeezed vacuum
states with the same squeezing parameters in the $(\hat{b}_1,
\hat{b}_3)$ representation may be transferred to a product
squeezed vacuum state of two squeezed vacuum states with the same
squeezing amplitudes but opposite phases in the $(\hat{B}_1,
\hat{B}_3)$ representation, while a product squeezed vacuum state
of two squeezed vacuum states with the same squeezing amplitudes
but opposite phases in the $(\hat{b}_1, \hat{b}_3)$ representation
is transferred to a two-mode squeezed state in the $(\hat{B}_1,
\hat{B}_3)$ representation. And a product squeezed vacuum state of
two squeezed vacuum states with the same squeezing parameters in
the $(\hat{B}_1, \hat{B}_3)$ representation is transferred to a
two-mode squeezed state with the same squeezing parameter in the
$(\hat{b}_1, \hat{b}_3)$. These transformation relations  are
explicitly expressed as
\begin{equation}
\label{56}
|\xi, \xi\rangle =|\xi, -\xi), \hspace{0.3cm}
|\xi, -\xi\rangle =|i\xi )_{B_1 B_3}, \hspace{0.3cm}
|\xi, \xi) = |\xi \rangle_{b_1 b_3}.
\end{equation}

In what follows we shall investigate generation of entangled squeezed vacuum states for the case
in which BECs are initially in the two product squeezed vacuum states in the $(\hat{b}_1, \hat{b}_3)$
representation $|\xi, -\xi\rangle$.

In this case, two BECs in the $(\hat{b}_1, \hat{b}_3)$ modes are
initially in a product squeezed vacuum state of two squeezed
vacuum states with the same squeezing amplitudes and the $\pi$
phase difference. From Eq. (\ref{16}) we know that after
transferring to the $(\hat{B}_1, \hat{B}_3)$ representation, the
system under our consideration is initially in a two-mode squeezed
vacuum state. This initial state can be explicitly written as
\begin{equation}
\label{57}
|\Phi(0)\rangle=\frac{1}{\cosh r}\sum^{\infty}_{n =0}
\left[-ie^{i\theta}\tanh r\right]^n |n, n),
\end{equation}
where $\xi=r\exp(i\theta)$, with $r$ and $\theta$ real and positive.
Then making use of  Eqs. (\ref{21}), (\ref{22}), and (\ref{57}) we know that at time $t$ the system will be a state
\begin{eqnarray}
\label{58}
|\Phi(t)\rangle&=&\frac{1}{\cosh r}\sum^{\infty}_{n=0}
\exp\left\{it\left[(q+\chi-2g)n-(3q+\chi)n^2\right]\right\} \nonumber \\
& &\times \left[-ie^{-i\theta }\tanh r\right]^n |n, n).
\end{eqnarray}

When relevant parameters satisfy the conditions $q =2\chi$ and
$4g=-19q$, the wavefunction of the system (\ref{58})  becomes
\begin{eqnarray}
\label{59}
|\Phi(\tau)\rangle&=&\frac{1}{\cosh r}\sum^{\infty}_{n=0}
\exp\left[-\frac{i}{2}\tau n(n-3)\right] \nonumber \\
& &\times \left[-ie^{-i\theta}\tanh r\right]^n |n, n),
\end{eqnarray}
where we have set $\tau=7qt$.

We note that the wavefunction of the system (\ref{59}) differs
from a conventional two-mode squeezed  state  (\ref{57}) by an
extra phase factor appearing in its decomposition into a
superposition of Fock states. It can always be represented as a
continuous sum of two-mode squeezed states. And under appropriate
periodic conditions, it can reduce to discrete superpositions of
two-mode squeezed states. It is this point that we use in present
paper to create entangled squeezed states what we expect.
Actually, the state (\ref{59}) can be expressed as a continuous
superposition of two-mode squeezed states
\begin{equation}
\label{60}
|\Phi(\tau)\rangle = \int^{2\pi}_{0}\frac{d\phi}{2\pi} g(\phi)
|ie^{i\phi} \xi )_{B_1B_3},
\end{equation}
where the phase $ g(\varphi)$ function is given by
\begin{equation}
\label{61} g(\phi)= \sum^{\infty}_{n=0}\exp\left[-i\frac{1}{2}\tau
n(n-3)-in\phi\right]
\end{equation}

Since $n(n-3)$ is always even, the exponential function
$\exp[-i\tau n(n-3)/2]$ in Eq. (\ref{59}) is periodic function
with the period $T=2\pi$. When $\tau=(M/N)2\pi$ with $N$ and $M$
being mutually prime integers, the phase function $ g(\phi)$ is a
periodic function with respect to $n$ with the period $2N$. Hence,
the wavefunction may be expressed as a discrete superposition
state of two-mode squeezed states
\begin{equation}
\label{62}
\left|\Phi\left(\tau=\frac{M}{N}2\pi\right)\right\rangle = \sum^{2N-1}_{n=0} c_{r}\left|ie^{i\varphi_r}\xi \right)_{B_1B_3},
\end{equation}
where the running phase is defined by
\begin{equation}
\label{63}
\varphi_r=\frac{\pi}{N}r, \hspace{0.5cm} (r, s=0,1,2, \cdots, 2N-1).
\end{equation}
The coefficients in Eq. (\ref{62}) are given by
\begin{equation}
\label{64}
 c_r=\frac{1}{(2N)^2}\sum^{2N-1}_{n=0}\exp\left\{-\frac{\pi
i}{N}\left[nr-Mn(n-3)\right]\right\}.
\end{equation}

We now give two nontrivial examples of entangled squeezed vacuum
states. The first one is the  case of $N=2$ and $M=1$, i.e.,
$\tau=\pi$ in Eq. (\ref{62}). In this case, from Eq.(\ref{64}) we
find that there exist only two nonzero $c$-coefficients
$c_{1}=c^*_3 = 1/\sqrt2\exp(i\pi/4)$, which leads to the following
superposition state of two two-mode squeezed states with the same
squeezing amplitudes but opposite phase in the $(\hat{B}_1,
\hat{B}_3)$  representation
\begin{equation}
\label{65} \left|\Phi\left(\tau=\pi \right)\right\rangle =
\frac{1}{\sqrt2}[ |-\xi )_{B_1B_3} - i|\xi )_{B_1B_3}],
\end{equation}
where we have discarded the common phase factor $\exp(-i\pi/4)$ on the right-hand side of above equation.

After transferring to the $(\hat{b}_1, \hat{b}_3)$ representation, we obtain an entangled state of
two product squeezed vacuum states
\begin{equation}
\label{66} \left|\Phi\left(\tau=\pi\right)\right\rangle =
\frac{1}{\sqrt2} [ |i\xi, -i\xi \rangle - i|-i\xi, i\xi \rangle].
\end{equation}
where we have used Eq. (\ref{56}).

As the last example of creating entangled squeezed vacuum for
atomic BECs, we consider the situation of $N=4$ and $M=1$, i.e.,
$\tau=\pi/2$ in Eq. (\ref{62}).  In this case, from Eq.(\ref{64})
we find that all nonzero $c$-coefficients are $c_{0}=c_4 =1/2$,
and $c_{2}=-c_6=1/2\exp(i\pi/4)$, which results in the following
superposition state of four two-mode squeezed states with the same
squeezing amplitudes but different phases in the $(\hat{B}_1,
\hat{B}_3)$  representation
\begin{eqnarray}
\label{67} \left|\Phi\left(\tau=\frac{\pi}{2} \right)\right\rangle
&=& -\frac{1}{2} e^{i\frac{\pi}{4}}\left[\left|
\xi\right)_{B_1B_3} - \left|-\xi\right)_{B_1B_3}\right]
\nonumber \\
& &+ \frac{1}{2}\left[ \left| i\xi\right)_{B_1B_3} + \left|
-i\xi\right)_{B_1B_3}\right].
\end{eqnarray}

After transferring to the $(\hat{b}_1, \hat{b}_3)$ representation,
we find that the resulting entangled state is given by
\begin{eqnarray}
\label{68} \left|\Phi\left(\tau=\frac{\pi}{2}\right)\right\rangle
&=& -\frac{1}{2} e^{i\frac{\pi}{4}}\left[ \left| -i\xi, i\xi
\right\rangle
- \left| i\xi, -i\xi \right\rangle \right] \nonumber \\
& &+ \frac{1}{2}\left[ \left| \xi, -\xi\right\rangle + \left|
-\xi, \xi\right\rangle\right],
\end{eqnarray}
which is an entangled state of four product squeezed vacuum states.

\section{Concluding Remarks}
We have presented a scheme for the generation of entangled
squeezed states for atomic BECs with the Raman-coupled
configuration. In the proposed scheme quantum entanglement is
created through laser-atom interactions and inter-atomic
interactions in the BECs. Under the large detuning and exact
two-photon resonant condition, the atomic field operators at the
upper level is adiabatically eliminated, the system becomes an
effective two-mode system. In this process the laser-atom
interactions are converted as an atomic effective tunnelling
interaction between two lower levels with the tunnelling coupling
strength to be determined by strengths of the laser-atom
interactions and the laser detunings. We have discussed how to
create two-state and  multi-state entangled squeezed states and
superposition states of two-mode squeezed states. When the initial
state of the two-mode system is a product squeezed state with the
same squeezing amplitudes and phases, superposition states of
two-mode squeezed states can be created, while when the initial
state of the system is a product squeezed state with the same
squeezing amplitudes but opposite phases, entangled squeezed
states can be generated. We have found that generation of
different entangled states are strongly manipulated by varying the
initial states of the system. Thus, one can create a variety of
entangled states by preparing different initial states.

In our scheme the essential requirements to achieve entangled
squeezed states include the exact two-photon resonance, the large
detunning of laser frequencies with respect to relevant atomic
transitions, and manipulation of strengths of laser-atom
interactions and inter-atomic weak nonlinear interactions. The
former two can be realized through adjusting frequencies of
lasers. Laser-atom interaction strengths can be changed through
controlling polarizations and intensities of lasers.  Finally,
inter-atomic nonlinear interactions can be manipulated through
changing atomic scattering lengths in BECs. Recent experiments on
Feshbach resonances in a Bose condensate \cite{ino} have indicated
that the scattering length of ultracold atoms can be altered
through Feshbach resonance. It is also worth noting that the Yale
group \cite{orz} has successfully produced squeezed-state atomic
BECs. These experimental advances together with current mature
detecting techniques for atomic BECs provide us with the
possibility to create and to observe experimentally entangled
squeezed states in atomic condensates. However, it should be
mentioned that  quantum entanglement discussed in present paper is
of particular nature: the entangled subsystems are not spatially
separated. This characteristic may limit its use. How to make use
of such kind of quantum entanglement as a resource to carry out
quantum information processing is an interesting topics for
further study.

 \acknowledgments

This work was supported in part the National Fundamental Research
Program (2001CB309310), the China NSF under Grant Nos. 90203018
and 10075018, the State Education Ministry of China, the
Educational Committee of Hunan Province,and the Innovation Funds
from Chinese Academy of Sciences.

\end{document}